\documentclass[12pt,times]{article}
 \usepackage[pdftex]{color,graphicx}
\usepackage{enumerate}
\usepackage{amsmath}
\usepackage{amssymb}
\usepackage{alltt}
\usepackage{setspace}
\usepackage{subfigure}
\usepackage{sidecap}
\usepackage{sectsty}
\usepackage{authblk}
\author[1]{Yoni BenTov\thanks{Current affiliation: Institute for Quantum Information and Matter, California Institute of Technology, Pasadena, CA 91125, USA}}
\author[1,2]{A. Zee}
\affil[1]{\small Department of Physics, University of California, Santa Barbara, CA 93106, USA}
\affil[2]{\small Kavli Institute for Theoretical Physics, University of California, Santa Barbara, CA 93106, USA}
\partfont{\large}
\sectionfont{\large}
\subsectionfont{\small}
\usepackage[left=2cm,top=2cm,right=3cm,head=1cm,foot=1cm]{geometry}
\usepackage{mathrsfs}
\numberwithin{equation}{section}
\let\oldsqrt\sqrt
\def\sqrt{\mathpalette\DHLhksqrt}
\def\DHLhksqrt#1#2{%
\setbox0=\hbox{$#1\oldsqrt{#2\,}$}\dimen0=\ht0
\advance\dimen0-0.2\ht0
\setbox2=\hbox{\vrule height\ht0 depth -\dimen0}%
{\box0\lower0.4pt\box2}}
\newcommand{\al}{\alpha}
\newcommand{\g}{\gamma}
\newcommand{\e}{\varepsilon}
\newcommand{\ta}{\theta}

\newcommand{\G}{\Gamma}
\newcommand{\Gs}{\mathscr G}
\newcommand{\Gc}{\mathcal G}
\newcommand{\ph}{\varphi}
\newcommand{\da}{\dagger}
\newcommand{\pa}{\partial}

\newcommand{\la}{\mathscr{L}}

\newcommand{\M}{\mathcal M}

\newcommand{\x}{\times}
\newcommand{\ox}{\otimes}

\newcommand{\ml}{\left(\begin{matrix}}
\newcommand{\mr}{\end{matrix}\right)}

\newcommand{\C}{\mathcal C}

\newcommand{\W}{\Omega}

\newcommand{\bra}{\langle}
\newcommand{\ket}{\rangle}
\newcommand{\tr}{\text{tr}}

\newcommand{\del}{\delta}
\newcommand{\Del}{\Delta}

\newcommand{\ep}{\epsilon}
\newcommand{\re}{\text{Re}}
\newcommand{\im}{\text{Im}}
\newcommand{\sgn}{\text{sgn}}
\newcommand{\half}{\tfrac{1}{2}}
\newcommand{\third}{\tfrac{1}{3}}

\newcommand{\sixth}{\tfrac{1}{6}}

\newcommand{\s}{\sigma}

\newcommand{\diag}{\text{diag}}
\newcommand{\sq}{\tfrac{1}{\sqrt2}}

\newcommand{\Ds}{\mathscr D}

\newcommand{\As}{\mathscr A}
\newcommand{\Bs}{\mathscr B}
\newcommand{\Cs}{\mathscr C}

\newcommand{\Hs}{\mathscr H}

\newcommand{\Ms}{\mathscr M}

\newcommand{\Ta}{\Theta}

\begin{document}
\title{The Origin of Families and $SO(18)$ Grand Unification}
\date{}
\maketitle
\abstract{We exploit a recent advance in the study of interacting topological superconductors to propose a solution to the family puzzle of particle physics in the context of $SO(18)$ (or more correctly, $Spin(18)$) grand unification. We argue that Yukawa couplings of intermediate strength may allow the mirror matter and extra families to decouple at arbitrarily high energies. As was clear from the existing literature, we have to go beyond the Higgs mechanism in order to solve the family puzzle. A pattern of symmetry breaking which results in the $SU(5)$ grand unified theory with horizontal or family symmetry $USp(4) = Spin(5)$ (or more loosely, $SO(5)$) leaves exactly three light families of matter and seems particularly appealing. We comment briefly on an alternative scheme involving discrete non-abelian family symmetries. In a few lengthy appendices we review some of the pertinent condensed matter theory.}
\section{The family problem}
Forces are unified, but matter is not. That quarks and leptons are repeated in three families is one of the most nagging puzzles in particle physics. Long ago, it was observed that the spinor representation of orthogonal groups, upon restriction to an orthogonal subgroup, decomposes into a bunch of spinors of the smaller group in a repetitive structure highly suggestive of the observed families \cite{gell-mann ramond slansky, wilczek zee, fujimoto, bagger dimopoulos}.\footnote{In addition to the highly attractive repetitive structure provided naturally by the theory of orthogonal groups, we also find it intriguing that spacetime is also governed by an orthogonal group, namely the
Lorentz group $SO(3,1)$ (or more correctly, $Spin(3,1) = SL(2,C)$).} In the $SO(18)$ (strictly, $Spin(18)$) grand unified theory (GUT), all known fermions are components of a single irreducible $256^+$ spinor representation, and matter is unified at high energy scales. 
\\\\
Unfortunately, in this scheme, in addition to the desired $16^{+}$s to which quarks and leptons belong in $SO(10)$ (strictly, $Spin(10)$)
unification \cite{GUT}, we also obtain an equal number of $16^{-}$s which are unknown experimentally. In the original literature, it was suggested that these $16^{-}$s acquire large masses and/or are permanently confined by a ``heavy color" gauge group. See the references cited above for details. 
\\\\
A seemingly unsurmountable stumbling block is that in the standard Higgs mechanism, large fermion masses necessarily break the gauge symmetry and give the gauge bosons large masses, contrary to observation. Thus, it was already clear long ago that, in order for spinorial unification to work, we must somehow evade or go beyond the Higgs mechanism. Recent developments in condensed matter physics afford us precisely this opportunity, which we will outline in detail below.
\section{Mass without mass terms: Kitaev-Wen mechanism}
It was recently argued by X. G. Wen that the $Spin(10)$ unified theory can be regularized on a $3d$ spatial lattice with continuous time \cite{wen spin(10)}. The low energy limit of lattice gauge theory is necessarily non-chiral \cite{fermion doubling}, so the continuum fermion fields that emerge from a $Spin(10)$ lattice gauge theory must transform as the reducible representation $16^+\oplus 16^-$. The conclusion of Wen's paper implies that the $16^-$ mirror fermions must have somehow obtained mass and decoupled from the low energy theory without breaking $Spin(10)$ and without giving mass to the gauge bosons. (Later, it was realized that the method of defect condensation can also be employed to support this conclusion \cite{cenke Z16, cenke GUT}. See Appendix~\ref{sec:defects}.)
\\\\
The same type of argument was independently proposed by A. Kitaev to show that the free-fermion classification of $^3$He-B reduces under interactions from $Z$ to $Z_{16}$ \cite{kitaev Z16}. This means that $16$ copies of topological superconductor\footnote{The reader who is unfamiliar with the concept of topological superconductor may wish to consult \cite{kitaev periodic table} and \cite{ludwig classification}. However, it is not necessary to master these references in order to understand our paper.} can be smoothly deformed into an ordinary superconductor without going through a bulk phase transition and without breaking time reversal invariance. (This conclusion was later verified by a different approach \cite{fidkowski chen vishwanath}.) This means the protected gapless $(2+1)$-dimensional edge modes decouple from the low energy theory even though time reversal symmetry forbids all mass terms in the Lagrangian. This also means that the $(3+1)$-dimensional bulk theory can be tuned through the point $m = 0$ without closing the bulk mass gap. Lattice simulations supporting these types of arguments have also appeared recently \cite{mass without condensates 1, mass without condensates 2}.\footnote{After the first draft of our paper was posted, additional numerical work appeared to further support the idea that this transition can be described by a continuum field theory \cite{catterall}.}
\\\\
Therefore, from these recent developments in condensed matter theory, we learn that in very special cases, one of which serendipitously happens to be the Standard Model (SM), it is possible for the fermion single particle spectrum to obtain an interaction-induced energy gap without any explicit fermion mass term in the Lagrangian. We will refer to this argument for ``mass without mass terms" as the Kitaev-Wen mechanism. We emphasize again that in this approach, in contrast to the Higgs mechanism, the electroweak gauge symmetry remains unbroken and the gauge bosons remain massless. (This point will be discussed in Sec.~\ref{sec:massless gauge bosons}.)
\\\\
Let us state at the outset that this argument forces us to carefully reconsider some long-standing wisdom about how fermions obtain mass, and it is certainly a radical departure from the standard orthodoxy in particle physics. We will follow closely the argument as presented in \cite{wen spin(10)}, focusing on those aspects which pertain to particle physics rather than those pertaining to the continuum limit of lattice gauge theory.
\subsection{Single particle spectrum in $Spin(10)$ unification}\label{sec:hamiltonian}
Let $\psi$ transform as a left-handed chiral spinor of the Lorentz group and as a $16^+$ chiral spinor of $Spin(10)$:
\begin{equation}
\psi \sim (2,1)\;\text{ of } Spin(3,1)\;,\;\; \psi \sim 16^+\;\text{ of } Spin(10)\;.
\end{equation}
We use the two-component Weyl spinor notation for the Lorentz group but the reducible Dirac spinor notation for the $Spin(10)$ flavor group. The field $\psi$ has $2\x 32 = 64$ components, half of which are set to zero by the condition $\half(I_{32\x 32}-\G_F)\psi = 0$, where $\G_F$ is the chirality matrix of $Spin(10)$.
\\\\
As is well known, since $16^+\ox_S 16^+ = [1]\oplus [5]^+$ (where $[k]$ denotes the $k$-index antisymmetric tensor, and the superscript $+$ on the right-hand side denotes self-duality) we can introduce scalar fields $\phi$ transforming as $[1]$ and $[5]^+$ and write down the following Yukawa interactions: 
\begin{equation}\label{eq:so(10) yukawas}
\la_{\text{Yuk}} = -\half\,\psi^T \!\left(\!\phantom{\frac{}{}} i\s_2\ox \left( y_{10}\,\phi_a C\G^a+y_{126}\,\phi_{a_1...a_5}C\G^{a_1}...\,\G^{a_5}\right)\right) \psi + h.c.
\end{equation}
Here $a = 1,...,10$ labels the vector representation of $Spin(10)$, and $\G^a$ and $C$ are the gamma matrices and charge conjugation matrix for $Spin(10)$, which are $32$-by-$32$ matrices. The 5-index tensor $\phi_{a_1...a_5}$ has an implied total antisymmetry and self-duality condition in $Spin(10)$ and hence has $\half \ml 10\\ 5 \mr = 126$ independent components. 
\\\\
Let us set $y_{126} = 0$ and only consider the coupling to $\phi_a \sim 10$. (We will briefly consider the $126$ in Appendix~\ref{sec:defects in 126}.) If this field condenses, say $\bra\phi_a\ket = v\,\del_{a,10}$, then the $Spin(10)$ theory breaks to $Spin(9)$, and all fermions obtain mass. (The $16^+$ and $16^-$ of $Spin(10)$ become the same $16$-spinor of $Spin(9)$, and a bare mass term would be allowed in a $Spin(9)$ theory.)
\\\\
The usual assumption is that when this field does not condense, i.e. $\bra\phi_a\ket = 0$, the fermions are massless. In fact this assumption is based on weak coupling perturbation theory and must be re-examined when the coupling is strong enough so as to invalidate a simple perturbative expansion. Moreover, strong coupling methods such as those employed in \cite{no chiral fermions} roughly correspond to an expansion around $y = \infty$, and those methods may have to be re-examined when $y$ is not so large. This means we are interested in the region of intermediate coupling \cite{wang wen}, heuristically meaning $y \sim 1/y$ or $y \sim 1$.
\\\\
To study the single particle dispersion relation in the regime of intermediate Yukawa coupling, we will pass to the Hamiltonian formalism and a Majorana description of the fermions. If $\psi$ is a left-handed Weyl spinor, then we can define a 4-component Majorana spinor\footnote{This satisfies $\Psi = \Psi^c$ with $\Psi^c \equiv i\g^2 \Psi^*$.} 
\begin{equation}\label{eq:majorana spinor}
\Psi \equiv \ml \psi\\ - i\s_2 \psi^* \mr\;.
\end{equation}
For further convenience, also define the matrix-valued fields
\begin{equation}
\Phi_R \equiv \half\{C,\G^a\}\phi_a = \half \sum_{i\,=\,1}^5\{C,\G^{2i}\}\phi_{2i}\;,\;\; \Phi_I \equiv \tfrac{1}{2i}[C,\G^a] = \tfrac{1}{2i}\sum_{i\,=\,1}^5[C,\G^{2i-1}]\phi_{2i-1}\;.
\end{equation}
The Hamiltonian density for a Majorana spinor with the Yukawa interaction in Eq.~(\ref{eq:so(10) yukawas}) with $y_{126} = 0$ is:
\begin{equation}
\Hs = -\half\overline{\Psi} \left\{ \g^i \ox I\;i\pa_i+y_{10}\left(I\ox \Phi_R-i\g^5\ox\Phi_I \right)\right\}\Psi\;.
\end{equation}
Let us analyze this Hamiltonian using a Born-Oppenheimer approximation in which we treat $\phi_a$ as a slowly varying background field. Then to leading order we can drop the gradients of $\phi_a$ and then Fourier transform to momentum space to obtain $\int d^3x\, \Hs = \int\frac{d^3p}{(2\pi)^3}\half \widetilde\Psi^\da(\vec p\,) \,H(\vec p,\vec\phi\,)\, \widetilde\Psi(\vec p\,)$, where
\begin{equation}
H(\vec p,\vec\phi\,) = \g^0\g^i\ox I\;p_i-y_{10}\left( \g^0\ox \Phi_R-i\g^0\g^5\ox \Phi_I\right)\;.
\end{equation}
Since $P \equiv \g^0\g^i\ox I p_i$ anticommutes with $M \equiv -y_{10}\left( \g^0\ox \Phi_R-i\g^0\g^5\ox \Phi_I\right)$, and since $\Phi_R^{\,2} = \left(\sum_{i\,=\,1}^5\phi_{2i}^{\,2}\right) I$, $\Phi_I^{\,2} = \left(\sum_{i\,=\,1}^5\phi_{2i-1}^{\,2}\right) I$, and $\{\Phi_R,\Phi_I\} = 0$, we can square the above Hamiltonian to obtain the single particle dispersion relation:
\begin{equation}\label{eq:dispersion}
E(\vec p,\vec\phi\,)^2 = \vec p^{\,2}+y_{10}^2\,\vec\phi^{\,2}\;.
\end{equation}
The energy required to produce a single fermion above the vacuum is then:
\begin{equation}\label{eq:gap}
\Del(\vec\phi) \equiv |E(\vec p = 0,\vec\phi)| = |y_{10}|(\vec\phi^{\,2})^{1/2}\;.
\end{equation}
In the path integral for $\phi_a$, the quantity $(\vec\phi^{\,2})^{1/2}$ in Eq.~(\ref{eq:gap}) should be understood as $\bra\vec\phi^{\,2}\ket^{1/2}$. If the field $\phi_a$ has zero mean then the $Spin(10)$ symmetry remains unbroken:
\begin{equation}\label{eq:condition 1}
\bra \vec\phi\ket = 0\qquad (\text{no symmetry breaking})\;.
\end{equation}
If the field $\phi_a$ satisfies Eq.~(\ref{eq:condition 1}), then the variance
\begin{equation}\label{eq:variance}
\s_\phi \equiv \left( \bra \vec\phi^{\,2}\ket-\bra\vec\phi\,\ket^2\right)^{1/2}
\end{equation}
simply equals $\bra\vec\phi^{\,2}\ket^{1/2}$, namely the quantity in Eq.~(\ref{eq:gap}). Therefore, if the symmetry is unbroken, it nevertheless costs nonzero energy to create a fermion provided that the field also satisfies:
\begin{equation}\label{eq:condition 2}
\bra \vec\phi^{\,2}\ket \neq 0\qquad (\text{nonzero gap})\;.
\end{equation}
If Eqs.~(\ref{eq:condition 1}) and~(\ref{eq:condition 2}) are satisfied simultaneously, the fermions appear to have mass without symmetry breaking. 
\\\\
The condition in Eq.~(\ref{eq:condition 2}) could be violated if the scalar field theory admits topological defects, since the Euclidean action would have a nonzero imaginary part\footnote{Consider the following toy example from ordinary statistics. Let $\phi\in(-\infty,+\infty)$ be a real random variable drawn from the gaussian distribution
\begin{equation}
P(\phi) = e^{-\half m^2\phi^2}\;,\;\; m^2 > 0\;.
\end{equation}
The partition function is $Z = \int_{-\infty}^\infty d\phi\, P(\phi)$, and the quantity
\begin{equation}
\bra\phi^2\ket = \frac{1}{Z}\int_{-\infty}^\infty \!\!\!\! d\phi\,P(\phi)\,\phi^2 = \frac{1}{m^2}
\end{equation}
is obviously positive. Now consider a modified distribution:
\begin{equation}
\widetilde P(\phi) \equiv P(\phi)\,e^{\,\half i\pi\, \sgn(\phi)}\;.
\end{equation}
Then we have:
\begin{equation}
\int_{-\infty}^\infty \!\!\!\!d\phi\, \widetilde P(\phi)\,\phi^2 = \int_0^{\infty}\!\!\!\!d\phi\, P(\phi)\,(+i)\,\phi^2+\int_{-\infty}^0\!\!\!\!d\phi\, P(\phi)\,(-i)\,\phi^2 = i\left( \int_0^{\infty}\!\!\!\!d\phi\,P(\phi)\,\phi^2-(-1)^2\int_0^{\infty}\!\!\!\!d\phi\,P(\phi)\,\phi^2\right) = 0\;.
\end{equation}
The modified partition function $\widetilde Z = \int_{-\infty}^\infty d\phi\,\widetilde P(\phi)$ itself vanishes, so strictly speaking the expectation values are indeterminate. For the bosonic path integral in Euclidean signature, if the action without topological terms is denoted $S_0$ and the topological contributions are denoted $iS_{\text{top}}$ (with $S_0$ and $S_{\text{top}}$ real), then the quantity $P(\phi)$ plays the role of $e^{-S_0}$, and the quantity $\widetilde P(\phi)$ plays the role of $e^{-S_0+iS_{\text{top}}}$.
}. However, for a fixed value of the $Spin(10)$-invariant magnitude $||\phi|| \equiv (\vec\phi^{\,2})^{1/2}$, the angular variables $\hat\phi_a \equiv \phi_a/||\phi||$ satisfy:
\begin{equation}
\hat\phi_a \in S^9\;\;.
\end{equation}
The topological charges of defects take values in the homotopy groups $\Pi_k(S^9)$ for $k = 0,...,D$, where $D$ is the number of physical spatial dimensions. Since 
\begin{equation}
\Pi_0(S^9) = \Pi_1(S^9) = \Pi_2(S^9) = \Pi_3(S^9) = 0
\end{equation}
there are no stable topological defects. Configurations for which $\bra \vec\phi^{\,2}\ket = 0$ contribute with measure zero to the scalar field path integral, and the quantity $\Del(\vec\phi)$ in Eq.~(\ref{eq:gap}) is nonzero.
\\\\
However, even when there are no defects and both conditions Eqs.~(\ref{eq:condition 1}) and~(\ref{eq:condition 2}) are satisfied, the effective action for $\phi_a$ may in principle contain a $\Ta$ term. Depending on the value of $\Ta$ the theory may still contain massless particles (see Appendix~\ref{sec:wzw}). Fortunately, we have
\begin{equation}
\Pi_4(S^9) = 0\;
\end{equation}
so there is no $\Ta$ term in the effective action for $\phi_a$. The single particle spectrum is indeed fully gapped as indicated by Eq.~(\ref{eq:gap}). 
\\\\
Furthermore, in 3+1 dimensions the quantity 
\begin{equation}
\bra\vec\phi^{\,2}\ket \equiv \lim_{y\to x}\bra\vec\phi(x)\vec\phi(y)\ket
\end{equation}
is a quantity of the order of magnitude of the energy cutoff in the theory. In this paper we will assume complete ignorance as to the appropriate resolution of the hierarchy problem, and therefore we treat cutoff-dependent quantities in scalar field theory as free parameters. So not only do the fermions have mass, but they have a mass which may in principle be much higher than the scale of $Spin(10)$ unification. 
\\\\
Note that this is consistent with the unitarity bounds obtained by Appelquist and Chanowitz \cite{appelquist}. The reason is that they introduce bare fermion mass terms which explicitly break the $SU(2)\x U(1)$ electroweak gauge symmetry. Their analysis is perfectly self-consistent, but it does not provide any bound on the single-particle gap generated via the Kitaev-Wen mechanism, in which explicit mass terms never appear, and $SU(2)\x U(1)$ remains unbroken.
\\\\
So far we have treated $Spin(10)$ as a global symmetry and have said nothing about the corresponding gauge theory. If this method is to have any relevance to unification in particle physics, we must explain why the gauge bosons remain massless while the fermions obtain mass.
\subsection{Massless gauge bosons}\label{sec:massless gauge bosons}
At this stage, the goal is to explain how to recover the GUT-scale phenomenology of the $Spin(10)$ unified theory: massless ordinary fermions, massless gauge bosons, and no mirror fermions. So if we are to embed the theory in $Spin(18)$, then the condition of ``no mirror fermions" means ``decouple the mirror fermions at energy scales above $M_{\text{GUT}} \sim 10^{16}$ GeV." 
\\\\
Concretely, for one family of matter $\psi \sim 16^+$ and one family of mirror matter $\psi' \sim 16^-$, we are interested in the following Lagrangian:
\begin{align}
\la &= \tfrac{1}{2} \tr\left( X_{\mu\nu} X^{\mu\nu}\right)+\half (D_\mu \phi)^T (D^\mu \phi) + \psi^\da \bar\s^\mu i D_\mu \psi + \psi'^\da \bar\s^\mu i D_\mu \psi' \nonumber\\
&- \half\sum_{a\,=\,1}^{10} \left(y_{10\,} \psi^T i\s_2 \ox C \G^a \psi + y_{10}' \,\psi'^T i\s_2 \ox C\G^a \psi' + h.c.\right)\phi_a\;.
\end{align}
Here $X_{\mu\nu} = \pa_\mu X_\nu - \pa_\nu X_\mu +\half g[X_\mu,X_\nu]$ is the $Spin(10)$ field strength, $X_\mu$ is the matrix of $Spin(10)$ gauge fields, $(D_\mu \phi)_a = (\del^{ab}\pa_\mu+g X_\mu^{ab}) \phi_b$ is the $Spin(10)$ gauge covariant derivative for the scalar field $\phi_a \sim 10$, and $D_\mu\psi = (I\pa_\mu +gX_\mu^{ab}(\half \Sigma_{ab}P_+))\psi$, $D_\mu\psi' = (I\pa_\mu+g X_\mu^{ab}(\half\Sigma_{ab}P_-))\psi'$ are the $Spin(10)$ covariant derivatives\footnote{Here $\Sigma_{ab} \equiv -i\half [\G_a,\G_b]$ and $P_\pm \equiv \half (I\pm \G_F)$.} for the fermions $\psi \sim 16^+$, $\psi' \sim 16^-$. 
\\\\
Let $T_i^{ab} = -T_i^{ba}$ ($i = 1,...,45$) generate the 10-representation of $Spin(10)$, and let $X_\mu^{ab} \equiv \sum_{i\,=\,1}^{45}X_\mu^i T_i^{ab}$. If we expand the kinetic term for the scalar field, we find the usual quadratic interaction for the gauge fields:
\begin{equation}
X^i_\mu \Ms^2_{ij} X^{j\mu}\;,\;\; \Ms^2_{ij} = g^2\, \phi_a (T_{i} T_{j})_{ab}\phi_b = g^{2} \sum_{a\,=\,1}^{10}\phi_a^{\,2}\;.
\end{equation}
If we naively compare this to Eq.~(\ref{eq:gap}), we might worry that this strong coupling argument also gives mass to the gauge bosons and hence does not reproduce the usual GUT-scale phenomenology. But the formula on its own may be misleading, and we have to be more careful when interpreting the interactions of the various fields with the scalar. To understand this, let us reconsider the difference between the ordinary fermions and the mirror fermions.
\\\\
Since the coupling of the ordinary fermions to the Higgs field is taken to be weak ($y_{10} \ll 1$), the ordinary fermions perceive the Higgs field as a collection of individual bosons which can be exchanged with an amplitude $y_{10}$. However, since the coupling of the mirror fermions to the Higgs field is \textit{not} taken to be weak ($y_{10}' \sim 1$), the mirror fermions see a wildly fluctuating scalar field instead of a collection of particles. Therefore, \textit{as far as the mirror fermions are concerned} we may replace the ``fluctuations" by the slowly-varying ``trend," and thereby drop the Higgs kinetic term and replace $\phi_a$ with a constant.
\\\\
The issue of whether the gauge bosons obtain mass then depends on whether the gauge coupling $g$ is weak. Since the usual scheme is to assume perturbative unification, we should have $g$ small: as far as the gauge bosons are concerned we cannot drop the Higgs kinetic term, and we have massless gauge bosons exchanging individual Higgs bosons as in the usual picture.
\subsection{Physical picture}\label{sec:picture}
The claim that, in certain special situations, fermions can obtain mass without mass terms in the Lagrangian is so counter to the standard orthodoxy in particle physics that we should at least attempt to provide some physical picture for what might be going on. One precise way to think about this phenomenon is to focus on the propagator for the fermion field. The Lehmann-Kallen decomposition for the fermion propagator is\footnote{We use metric signature $\eta_{\mu\nu} = \diag(+,-,-,-)$.}:
\begin{equation}\label{eq:lehmann}
-i\Ds(p) = \int_0^{\infty}ds\;\frac{ \rho_1(s)\not\! p+s^{1/2}\rho_2(s)I}{p^2-s+i\e}\;.
\end{equation}
The spectral density functions $\rho_1(s)$ and $\rho_2(s)$ are constrained by positivity to satisfy the following inequalities \cite{IZ}:
\begin{equation}\label{eq:positivity}
\rho_1(s) \geq 0\;,\;\; \rho_1(s) \geq |\rho_2(s)|\;.
\end{equation}
Consider the transformation $\psi \to \g^5\psi$ that flips the sign of the mass bilinear $\bar\psi\psi$. Under this transformation, the function $\rho_1(s)$ is even while the function $\rho_2(s)$ is odd. If this transformation is a symmetry of the effective Lagrangian, then $\rho_2(s)$ must be zero. If there is an isolated single particle pole at $s = \Del^2$ with residue 1, then the propagator for a theory of ``massive fermions without mass terms" takes the form:
\begin{equation}\label{eq:rho2=0}
-i\Ds(p) = \left(\frac{1}{p^2-\Del^2+i\e}+\int_{m_{\text{th}}}^\infty ds\;\frac{\s(s)}{p^2-s+i\e}\right)\not\! p
\end{equation}
where $\s(s) \equiv \rho_1(s)-\del(s-\Del^2)$ is the spectral density without the single-particle pole, and the scale $m_{\text{th}}$ determines the onset of the multiparticle continuum. Note that, while unfamiliar, the condition $\rho_2(s) = 0$ is perfectly consistent with the constraints in Eq.~(\ref{eq:positivity}).
\\\\
The rank of the matrix $\Ms \equiv \rho_1(s)\g^\mu p_\mu$ is double the rank of the matrix $\Ms' \equiv \rho_1(s) \g^\mu p_\mu+s^{1/2}\rho_2(s)I$, so one physical interpretation of the Kitaev-Wen mechanism is that the interactions generate a new ``soliton"-like sector of the theory with the same quantum numbers as the original particles. In this sense, we may think of the Kitaev-Wen mechanism as a ``particle doubling" effect \cite{bentov KF} that only becomes possible when the number of chiral fermions is a multiple of 16 (in 3+1 dimensions). 
\\\\
We will now argue that the Kitaev-Wen mechanism from condensed matter physics and lattice gauge theory allows us to propose a novel solution to the family puzzle in the context of the $Spin(18)$ theory of family unification.
\section{Family unification with $Spin(18)$} 
The usual symmetry breaking pattern for the $Spin(18)$ model is $Spin(18) \to Spin(10)\x Spin(8)$. Another potentially interesting pattern is the breaking to the maximal unitary subgroup, i.e. $Spin(18) \to U(9)$. We will explore both possibilities.
\subsection{$Spin(18) \to Spin(10)\x Spin(8)$}
The fermions will be denoted by $\Psi \sim 256^+$. The Higgs fields which can couple to the fermion mass bilinears in this model transform as $[1] = 18$, $[5] = 8568$, and $[9]^+ = 24310$. Consider the smallest Higgs representation and introduce a scalar field $\Phi_M \sim [1]$, where $M = 1,...,18$ labels the $Spin(18)$ vector:
\begin{equation}
\la_{\text{Yuk}} = Y_{18}\,\sum_{M\,=\,1}^{18}\Phi_M\,\Psi^T  i\s_2\ox \mathbb{C}\mathbb{G}^M\Psi+h.c.
\end{equation}
Here $\mathbb G^M$ and $\mathbb C$ are the gamma matrices and charge conjugation matrix for $Spin(18)$, which are $512$-by-$512$ matrices. Suppose $Spin(18)$ gets broken to $Spin(10)\x Spin(8)$ at a scale $M_{\text{UV}}$. The fermions break up into
\begin{align}
&\psi \sim (16^+,8^+)\;,\;\; \psi' \sim (16^-,8^-)
\end{align}
and the Higgs field breaks up into
\begin{align}
&\phi_a \sim (10,1)\;,\;\; \ph_i \sim (1,8_v)\;.
\end{align}
At energy scales below $M_{\text{UV}}$, we end up with Yukawa interactions of the form:
\begin{align}
\la_{\text{Yuk}} &= \sum_{a\,=\,1}^{10}\phi_a\left(y_{10}\,\psi^T i\s_2 \ox C\G^a \ox \C\,\psi+y_{10}'\,\psi'^{\,T} i\s_2 \ox C\G^a \ox \C\, \psi'\right)\nonumber\\
&+y_8\,\sum_{i\,=\,1}^8\ph_i\left( \psi^T i\s_2 \ox C \ox \C\Gc^i\,\psi'+\psi'^{\,T} i\s_2 \ox C \ox \C\Gc^i\,\psi\right)+h.c.
\end{align}
with the couplings $y_{10}$, $y_{10}'$, and $y_8$ no longer being equal. (As before, $\G^a$ and $C$ denote the gamma matrices and charge conjugation matrix for $Spin(10)$, and we have introduced $\Gc^i$ and $\C$ as the corresponding matrices for $Spin(8)$.)
\\\\
Suppose for the moment that the field $\ph_i \sim (1,8)$ plays no essential role. Then we have 8 copies of the situation described previously plus 8 copies of the same situation for mirror matter. Let $M_{\text{IR}}$ be an intermediate energy scale far below $M_{\text{UV}}$ but far above $v \sim 246$ GeV at which $SU(2)\x U(1)$ is broken:
\begin{equation}
v \ll M_{\text{IR}} \ll M_{\text{UV}}\;.
\end{equation}
The mirror matter will decouple via the Kitaev-Wen mechanism while the ordinary matter will remain massless provided that the mass-squared parameter $M_{\phi}^2$ for the $\phi_a \sim 10$ is positive and the Yukawa couplings satisfy:
\begin{equation}\label{eq:goal for yukawas}
y_{10}(M_{\text{UV}}) = y_{10}'(M_{\text{UV}}) \equiv Y_{18}(M_{\text{UV}})\;,\;\; y_{10}(M_{\text{IR}}) \ll y_{10}'(M_{\text{IR}}) \sim 1\;.
\end{equation}
The success of this approach rests completely on the dynamical plausibility of the conditions in Eq.~(\ref{eq:goal for yukawas}). Since we are interested in a strong coupling effect, it is difficult to say more about this issue.
\subsection{$Spin(18) \to U(9)$}
Now we will study an alternative symmetry breaking pattern based on $Spin(2n) \to U(n)$. For the case at hand $(n = 9)$, the positive-chirality spinor decomposes as: 
\begin{equation}\label{eq:spin(18) to u(9)}
256^+ \to [0]\oplus [2]\oplus [4]\oplus [6]\oplus [8] = 1\oplus 36\oplus 126\oplus 84^*\oplus 9^*\;.
\end{equation}
The equality denotes the dimension of the representation in $SU(9)$, which has a 9-index invariant epsilon symbol that can be used to raise and lower indices. It is convenient to think in terms of the familiar $SU(5)$ GUT, so we will organize the discussion in terms of the subgroup $SU(5)\x SU(4)$ of $SU(9)$ defined by breaking the fundamental representation in the obvious way:
\begin{equation}
9 \to (5,1)\oplus (1,4)\;.
\end{equation} 
To further organize the discussion, it is useful to introduce the notation
\begin{equation}
(a,b)_{[k]}\;.
\end{equation}
This denotes a representation ``$a$" of $SU(5)$, a representation ``$b$" of $SU(4)$, and the representation $[k]$ of $SU(9)$ in which the $(a,b)$ of $SU(5)\x SU(4)$ is contained.\footnote{This notation also reminds us that the $U(1)$ charge in $U(9) = (SU(9)/Z_9)\x U(1)$ of the representation $[k]$ is simply $k$.} In this notation, the matter and mirror matter of the $Spin(18)$ GUT transform as:
\begin{itemize}
\item Matter:
\begin{align}
&(5^*,1)_{[8]}\;,\;\; (5^*,1)_{[4]}\;,\;\; (5^*,6)_{[6]}\;,\;\;(10,1)_{[2]}\;,\;\;(10,1)_{[6]}\;,\;\; (10,6)_{[4]}\;.
\end{align}
\item Mirror matter:
\begin{align}
&(5,4)_{[4]}\;,\;\; (5,4^*)_{[2]}\;,\;\;(10^*,4)_{[4]}\;,\;\; (10^*,4^*)_{[6]}\;.
\end{align}
\end{itemize}
The decomposition in Eq.~(\ref{eq:spin(18) to u(9)}) also results in 16 SM-singlet antineutrinos, as it must. The representations which are invariant under $SU(5)$ but transform nontrivially under $SU(4)$ are $(1,6)_{[2]},(1,4)_{[6]}$, and $(1,4^*)_{[8]}$. The representations which are fully invariant under $SU(5)\x SU(4)$ are $(1,1)_{[0]}$ and $(1,1)_{[4]}$.
\\\\
Notice that the matter comes in real representations of $SU(4)$ while the mirror matter comes in the vectorlike combination $4\oplus 4^*$. Therefore, all $SU(4)$ anomalies cancel for the matter and mirror matter separately, and anomaly matching does not impose an obstacle for decoupling the mirror matter.
\\\\
Recall that in $SU(5)$ the mass terms come from the products $5^*\ox 10$ and $10\ox_A 10$, which couple to a $5^*$ and $5$ of Higgs, respectively. In $SU(4)$, we have $4\ox 4^* = 1\oplus 15_{\text{adjoint}}$, so we can write down the following Yukawa interactions for the mirror fermions  $\psi' \sim (5\oplus 10^*,4\oplus 4^*)$ and a Higgs field $H_i \sim 5$:
\begin{align}
\la_{\text{Yuk}} &= H_i \left( y_D\, \psi_{j\al}^{'T} i\s_2\, \psi^{'[ij]\al}+y_D'\, \psi_j^{'T\al}i\s_2\, \psi^{'[ij]}_\al\right)+H^{\da i}\ep_{ijk\ell m} \left( y_U\,\psi^{'T[jk]}_\al i\s_2\,\psi^{'[\ell m]\al}\right)+h.c.
\end{align}
Here $i,j = 1,...,5$ label the fundamental of $SU(5)$, and $\al = 1,...,4$ labels the fundamental of $SU(4)$. Since $H$ is a 5-component complex vector, we can define real fields $\chi_{2i-1} \equiv \re(H_i)$ and $\chi_{2i} \equiv \im(H_i)$ and observe that $H^\da H = \sum_{I\,=\,1}^{10}\chi_I^{\,2}$ is actually invariant under $SO(10)$ transformations. So the Kitaev-Wen argument in this case is exactly the same as before:
\begin{equation}
SU(5)/SU(4) = SO(10)/SO(9) = S^9\;,\;\; \Pi_k(S^9) = 0\;,\;\; k = 0,...,4\;.
\end{equation}
If $M_{\Phi}^2$ is positive, and if $y_D'$ and $y_U'$ are not perturbatively small, then the mirror fermions can obtain UV-scale masses and decouple at low energy.
\section{Three families with $SU(5)\x USp(4)$ symmetry}
We have argued that the mirror matter can decouple if certain conditions such as Eq.~(\ref{eq:goal for yukawas}) for the Yukawa couplings are satisfied at energies below the $Spin(18)$ unification scale. To reproduce only the experimentally observed families, we want the extra families of ordinary matter to decouple from the low energy theory as well. For this purpose we will explore a further symmetry reduction of the horizontal gauge group. (In this section we will continue to imagine that the symmetry reduction occurs via $Spin(18) \to U(9) \to SU(5)\x SU(4)$.)
\\\\
The breaking of $SU(4) = Spin(6)$ into $USp(4) = Spin(5)$ can be understood as the breaking of $SO(6)$ into $SO(5)$ defined by leaving one component of the $6$-vector fixed:
\begin{equation}
6 \to 5\oplus 1\;.
\end{equation}
The $4$ and $4^*$ then become two copies of the same $4$-component irreducible Dirac spinor of $Spin(5)$. The matter and mirror matter then transform as:
\begin{equation}
\underbrace{(5^*\oplus 10,1\oplus 1\oplus 1)}_{\text{known matter}}\oplus\underbrace{ (5^*\oplus 10, 5)}_{\text{extra matter}}\oplus\underbrace{(5\oplus 10^*,4\oplus 4)}_{\text{mirror matter}}
\end{equation}
Thus, as observed in Ref.~\cite{gell-mann ramond slansky, wilczek zee, fujimoto}, we are left with three families of matter which transform trivially under the horizontal gauge group. 
\\\\
If $USp(4)$ were to remain asymptotically free and induce an $SU(5)$-breaking fermion bilinear condensate, we would need to explain how the five extra families of $5^*\oplus 10$ could have masses much larger than 1 TeV while the known fermions have their experimentally measured masses. This was the original problem with the ``heavy color" idea.
\\\\
However, if there is a sufficiently large quantity of matter such that the $USp(4)$ gauge group is not asymptotically free, then we can use the Kitaev-Wen argument again, this time to decouple the extra families of ordinary matter. If we simply posit that the Yukawa couplings for the extra matter are also large, then the interactions with an appropriate scalar field (with positive mass-squared parameter) would decouple these fermions as well.
\section{$SU(5)\x SU(2)$ and non-abelian discrete groups}\label{sec:discrete symmetry}
Since we no longer require the confinement of heavy color to conceal the extra families, we could take a different point of view regarding the breaking of the $SU(4)$ horizontal gauge group\footnote{C. Luhn has studied in detail the breaking of an $SU(3)$ horizontal gauge group into non-abelian discrete subgroups \cite{luhn}.} and instead suppose that the \textit{nontrivial} representations describe the known families  \cite{georgi theory of flavor}. Then one could hope that the quark and neutrino mixing matrices could be explained by group theory \cite{theory of flavor, pakvasa sugawara, wilczek zee cabibbo, sartori flavor}.
\\\\
For example, let us continue the train of thought that lead to $USp(4)$ and further break the horizontal group down to $Spin(3) = SU(2)$. Then we have $5 \to 3\oplus 1\oplus 1$, so we get:
\begin{equation}
\underbrace{(5^*\oplus 10, 1\oplus 1\oplus 1 \oplus 1 \oplus 1 \oplus 1)}_{\text{extra matter}}\oplus \underbrace{(5^*\oplus 10,3)}_{\text{known matter}}
\end{equation}
where we have accordingly switched the identification of extra matter and known matter. It is then possible to conceive of a Higgs sector which would break this down to a discrete non-abelian subgroup of $SO(3)$ \cite{isotropy subgroups of so(3), break so(3) to discrete}. For example, in the breaking $SO(3) \to A_4$, the three families would transform as an irreducible triplet \cite{ma A4}. 
\\\\
However, part of the attraction of the group $A_4$ is its two nontrivial singlet representations, the $1'$ and $1''$. In $SO(3) \to A_4$, these come from the traceless symmetric tensor, $5_S \to 1'\oplus 1''\oplus 3$. (For a review, see Ref.~\cite{zee A4}.) In our scheme based on $SU(9)$ embedded in $Spin(18)$, the fermions will always transform as antisymmetric tensor representations $[k]$ of $SU(9)$, which will only give us spinors and vectors of $SU(4) = Spin(6)$, $USp(4) = Spin(5)$, $SU(2)\x SU(2) = Spin(4)$, or $SU(2) = Spin(3)$. 
\\\\
Along these lines, if we began with the $256^-$ spinor of $Spin(18)$ instead of the $256^+$, then the resulting representations of $SU(4)$ would be swapped: the matter would transform as $4\oplus 4^*$, and the mirror matter would transform as $1\oplus 1\oplus 6$. Under $SU(4) \to SU(2)$, we would have the $4$ and $4^*$ each breaking up into $2\oplus 2$. 
\\\\
Then we could conceive of breaking $SU(2)$ down to a discrete non-abelian subgroup, such as the double cover of $A_4$, which we call $A_4'$ \cite{double T}. However, in this particular scheme one would have to deal with at least four families since all of the matter fields would transform as doublets. Furthermore, the interesting representations $2'$ and $2''$  of $A_4'$ would come from the spin-$\frac{3}{2}$ representation of $SU(2)$ via $4 \to 2'\oplus 2''$ (for a review, see Ref.~\cite{zee A4'}) and hence could not come from $SU(9)$ unification. 
\section{Discussion}
We have argued that in principle it is possible for the mirror matter in $Spin(18)$ unification to completely decouple from the effective field theory at scales far above the weak scale without breaking $SU(2)\x U(1)$. The extra ordinary matter beyond the three known families can also decouple by the same argument. The main open issue is to determine convincingly that conditions such as Eq.~(\ref{eq:goal for yukawas}) can be satisfied in models of this kind. 
\\\\
Since the parameter $M_\phi^2$ in the Higgs potential is taken positive (remember that we do not want $\phi_a$ to condense), we can integrate out $\phi_a$ in Eq.~(\ref{eq:so(10) yukawas}) and obtain a low-energy effective 4-fermion interaction:
\begin{equation}
\la_{\text{eff}} = \psi^\da \bar\s^\mu i\pa_\mu \psi+\frac{y_{10}^{\,2}}{M_{\phi}^2}(\psi^T i\s_2\ox C\G^a\psi + h.c.)(\psi^T i\s_2 \ox C\G_a \psi + h.c.)\;.
\end{equation}
If perturbation theory is applicable, then the fermions are clearly massless since there is no mass term in the Lagrangian, and the 4-fermion interaction is an irrelevant operator. But if perturbation theory in $y$ is not applicable, then we cannot drop this operator in the low-energy theory. 
\\\\
On the other hand, if the coupling is not too large, then we cannot directly apply strong-coupling methods such as those of \cite{no chiral fermions}, and we might expect that the dynamics are not so strong as to generate an expectation value for the bilinear $\psi^T i\s_2\ox C\G^a\psi$. In this case, the fermion single particle spectrum should have a gap $\Del$ given by Eq.~(\ref{eq:gap}), which could be pushed up arbitrarily high above the usual scale of unification, while the $Spin(10)$ symmetry remains unbroken.
\\\\
Therefore, an alternative way to view the Kitaev-Wen argument is to say that the 4-fermion interaction generates mass without a mass term if $|y_{10}|$ is large enough such that perturbation theory is not applicable but not so large that dynamical symmetry breaking occurs. The situation is summarized as follows:
\begin{itemize}
\item $y_{10} \ll 1$: massless fermions exchanging scalar particles
\item $y_{10} \sim 1$: massive fermions with $Spin(10)$ invariance and hence without mass terms
\item $y_{10} \gg 1$: dynamical symmetry breaking $Spin(10) \to Spin(9)$ and massive fermions
\end{itemize} 
Once the extra matter has decoupled, the phenomenology becomes that of the usual $Spin(10)$ or $SU(5)$ unified models. We have intentionally emphasized only the situation in which none of the extra fermions have any influence below the usual scale $M_{\text{GUT}} \sim 10^{16}$ GeV, but this was just the simplest choice. We invite the interested reader to re-evaluate the possible importance of the additional states for TeV-scale physics.
\\\\
Given the phenomenological success of the Higgs mechanism in particle physics, one could ask whether the Kitaev-Wen mechanism could also do the job without spontaneous symmetry breaking. (We have already explained in Sec.~\ref{sec:massless gauge bosons} that this mechanism will not give mass to the gauge bosons, so this question pertains only to the fermion mass.) 
\\\\
The physical intuition from Sec.~\ref{sec:picture} implies that a fermion whose mass comes from the Kitaev-Wen mechanism has a propagator of the form
\begin{equation}
\frac{\not\!p}{p^{\,2}-\Del^2} = \frac{1}{p^{\,2}-\Del^2}\;\half\left[ (\not\! p+\Del)+(\not\! p -\Del)\right]
\end{equation}
below the multiparticle threshold. This expresses the physical distinction between a fermion mass obtained from the Higgs mechanism and one obtained from the Kitaev-Wen mechanism. It is still unclear what the full phenomenological implications would be for an alternative version of particle physics based on this mechanism for generating fermion masses.
\begin{center} \textit{Acknowledgements} \end{center}\label{sec:acknowledgements}
This work was supported by NSF grant PHY13-16748. Y. BenTov would like to thank A. Kitaev, E. G. Moon, A. Rasmussen, C. Xu, Y. Z. You, S. Catterall, and J. Preskill  for helpful discussions.
\appendix\label{sec:appendix}
\section{Condensation of topological defects}\label{sec:defects}
According to Eqs.~(\ref{eq:dispersion}) and~(\ref{eq:gap}), it costs an energy $\Del(\vec \phi)$ to create a single fermion above the vacuum. As long as $\bra\phi_a\ket = 0$ the $Spin(10)$ symmetry is unbroken, and as long as $\bra \vec\phi^{\,2}\ket \neq 0$, the energy gap $\Del(\vec\phi)$ remains nonzero. [Recall Eqs.~(\ref{eq:condition 1}) and~(\ref{eq:condition 2}).] If there are topologically nontrivial configurations of $\phi_a$, then the condition $\bra\vec\phi^{\,2}\ket \neq 0$ is violated. Since $\Pi_k(S^9) = 0$ for $k = 0,1,2,3$, the $Spin(10)$ theory with a 10-vector Higgs field does not admit stable topological defects, and the required conditions are satisfied. 
\\\\
As an alternative argument for decoupling the mirror matter in $Spin(18)$ unification at scales far above the electroweak breaking scale $v \sim 246$ GeV, it is enlightening to consider a situation with a smaller symmetry group than $Spin(10)$ in which there \textit{are} topological defects. 
\\\\
Then we could ask whether it is possible to Higgs the theory, generate fermion masses, and then restore the electroweak subgroup without closing the fermion gap by condensing the operator which creates topological defects.\footnote{For example, in the Ising model, one can imagine the disordered phase as the phase in which the kink operators have a nonzero vacuum expectation value.} Then the fermions would have mass without mass terms even in a theory which violates Eq.~(\ref{eq:condition 2}). Since generating mass without mass terms is the main underlying theoretical difficulty, we then expect that the symmetry can be enlarged to $Spin(10)$ without any dynamically induced spontaneous symmetry breaking.
\\\\
This approach to regularizing the SM on a $3d$ spatial lattice was first proposed by BenTov, You, and Xu \cite{cenke Z16} (motivated by the arguments of Wang and Senthil \cite{wang senthil}) and then carried out to completion by You and Xu \cite{cenke GUT} for the Pati-Salam (PS) model \cite{pati salam} with gauge group\begin{equation}
G_{\text{PS}} = Spin(6)\x Spin(4) = SU(4)\x (SU(2)_{\text{L}}\x SU(2)_{\text{R}})\;.
\end{equation}
Under the breaking $Spin(10) \to G_{\text{PS}}$, the matter fields break up as\footnote{
Under the breaking of $G_{\text{PS}}$ to the SM gauge group
\begin{equation}
G_{\text{SM}} = SU(3)_{\text{color}}\x SU(2)_{\text{weak}}\x U(1)_{\text{hypercharge}}
\end{equation}
the matter fields further break up as:
\begin{align}
&(4,2,1) \to (3,2,+\sixth)\oplus (1,2,-\half)\;,\nonumber\\
&(4^*,1,2) \to (3^*,1,+\third)\oplus (3^*,1,-\tfrac{2}{3})\oplus (1,1,+1)\oplus(1,1,0)\;.
\end{align}
}:
\begin{equation}
16^+ \to (4^+,2^+)\oplus (4^-,2^-) = (4,2,1)\oplus (4^*,1,2)\;.
\end{equation}
In the first expression on the right we use the notation appropriate for $Spin(6)\x Spin(4)$, and in the second expression we use the $SU(4)$ notation $4^+ \equiv 4$, $4^- \equiv 4^*$, and the $SU(2)_{\text{L}}\x SU(2)_{\text{R}}$ notation $2^+ \equiv (2,1)$, $2^- \equiv (1,2)$. Similarly, for the mirror matter we have:
\begin{equation}
16^- \to (4^+,2^-)\oplus (4^-,2^+) = (4,1,2)\oplus (4^*,2,1)\;.
\end{equation} 
We will use the $SU(4)\x SU(2)_{\text{L}}\x SU(2)_{\text{R}}$ notation and express all fermions as left-handed Weyl spinors of the Lorentz group with the following gauge group indices:
\begin{align}\label{eq:PS fermions}
\psi_{A\al} \sim (4,2,1)\;,\;\; \bar\psi^{A\dot\al} \sim (4^*,1,2)\;,\;\;A = 1,...,4\;,\;\; \al = 1,2\;,\;\; \dot\al = 1,2\;.
\end{align}
Here the bar is part of the name of the field and does not denote any sort of conjugation. Remember that, in contrast to the Lorentz group\footnote{Here the symbol ``$\simeq$" denotes only local equivalence.} $SL(2,C) \simeq SU(2)\x SU(2)$, here the two $SU(2)$ groups are self-conjugate, so hermitian conjugation raises and lowers dotted and undotted indices instead of exchanging them. 
\\\\
All mass terms in $3+1$ dimensions are forbidden by $G_{\text{PS}}$ invariance. The Dirac-type fermion bilinear transforms as
\begin{equation}
\psi_{A\al}^Ti\s_2 \bar\psi^{B\dot\beta} \sim (1\oplus 15_{\text{adj}},2,2)\;.
\end{equation}
Since the $(2,2)$ representation of $SU(2)_{\text{L}}\x SU(2)_{\text{R}}$ is the 4-vector representation of $Spin(4)$, we can introduce a Higgs field
\begin{equation}\label{eq:4-vector higgs}
\phi^m = (\phi_1,...,\phi_4) \equiv \s^m_{\al\dot\al}\Phi^{\dot\al\al} \sim (1,2,2)\;,\;\; \s^m_{\al\dot\al} \equiv (i\vec\s,I)\;.
\end{equation}
If this Higgs were to condense, then it would break $Spin(4) \to Spin(3)$ and give all fermions a mass through the following Yukawa interaction:
\begin{equation}
\la_{\text{Yuk}} = -y\,\psi^T_{A\al}i\s_2\,\e^{\al\beta}\s^m_{\beta \dot\beta}\bar\psi^{A\dot\beta} \phi_m+h.c.
\end{equation}
The two Majorana-type fermion bilinears transform as:
\begin{equation}
\psi_{A\al}^Ti\s_2 \psi_{B\beta} \sim (10^+,3,1)\oplus (6,1,1)\;,\;\; \bar\psi^{A\dot\al\,T}i\s_2 \bar\psi^{B\dot\beta} \sim (10^-,1,3)\oplus (6,1,1)
\end{equation}
where in $Spin(6)$ language, the $10^+$ is the self-dual 3-form, the $10^-$ is the anti-self-dual 3-form, and the 6 is the vector.
\\\\
The fermion content in Eq.~(\ref{eq:PS fermions}) is of course chiral, which is why until the arguments of Kitaev and Wen it was not known how to regularize the SM on a purely $3d$ lattice. However, Kaplan \cite{kaplan} showed that it is possible to obtain this chiral theory as the boundary of a non-chiral $(4+1)$-dimensional topological superconductor. (This is a generalization of the original Jackiw-Rebbi calculation \cite{jackiw rebbi}.) It is this setup in which the method of defect condensation can be used to provide independent support for the validity of the Kitaev-Wen mechanism \cite{cenke Z16, cenke GUT}. 
\\\\
We will now review this argument. Consider a $(4+1)$-dimensional spacetime with the gauge group $G_{\text{PS}}$. In $4+1$ dimensions the Lorentz group is $Spin(4,1)$, and the 4-component Dirac spinor is \textit{irreducible}. This means we need to augment the matter content in Eq.~(\ref{eq:PS fermions}) with a collection of mirror fermions:
\begin{equation}\label{eq:PS mirror fermions}
\psi^{'A}_\al \sim (4^*,2,1)\;,\;\; \bar\psi_A^{\,'\,\dot\al} \sim (4,1,2)\;.
\end{equation}
Both $\psi'$ and $\bar\psi'$ are also written as left-handed Weyl spinors in $(3+1)$-dimensional notation. In $(4+1)$-dimensional notation, $\psi$ and $\psi'$ form an irreducible Dirac spinor, and $\bar\psi$ and $\bar\psi'$ form another irreducible Dirac spinor\footnote{Here we have raised and lowered the conjugated $SU(2)$ indices with the invariant antisymmetric symbol: $\psi^{'*}_{A\al} \equiv \e_{\al\beta}\psi^{'*\beta}$, $\bar\psi^{'*\,A\dot\al} \equiv \e^{\dot\al\dot\beta}\bar\psi^{'*\,A}_{\dot\beta}$.}:
\begin{equation}\label{eq:4+1 spinors}
\Psi^{(1)}_{A\al} \equiv \ml \psi_{A\al}\\ -i\s_2\, \psi^{'*}_{A\al} \mr \sim (4,2,1)\;,\;\; \Psi^{(2)\,A\dot\al} \equiv \ml \bar\psi^{A\dot\al}\\ -i\s_2\,\bar\psi^{'*\,A\dot\al} \mr \sim (4^*,1,2)\;.
\end{equation}
 Since the $(4+1)$-dimensional theory is not chiral, a $G_{\text{PS}}$-invariant Dirac mass with a domain wall profile can be written down for both of these spinors:
\begin{equation}
\la_{\text{mass}}^{\,4+1} = -m(x^4)\left( \overline{\Psi^{(1)}}^{A\al}\Psi^{(1)}_{A\al}+\overline{\Psi^{(2)}}_{A\dot\al}\Psi^{(2)\,A\dot\al}\right)\;,\;\; m(x^4) = \left\{ \begin{matrix} -m\;,\;\; x^4 > 0\\ 0\;,\;\; x^4 = 0\\ +m\;,\;\; x^4 < 0 \end{matrix}\right.\;.
\end{equation}
The constant $m$ is assumed positive. For one of these fermions, say $\Psi \equiv \Psi^{(1)}$, the equation of motion possesses the solution:
\begin{equation}
\Psi = \xi(x^0,x^1,x^2,x^3)\,e^{-m|x^4|}\;,\;\; \sum_{\mu\,=\,0}^3i\g^\mu \pa_\mu \xi = 0\;,\;\; \half (I+\g^5)\xi = 0\;.
\end{equation}
Therefore, the $(3+1)$-dimensional interface at $x^4 = 0$ contains only the massless chiral fermions given in Eq.~(\ref{eq:PS fermions}). If the phase with $m > 0$ is the trivial gapped phase, then these fermions should be thought of as living on the boundary of the $m < 0$ ``topological" gapped phase.\footnote{Let us remind the reader what the word ``trivial" means in this context. An operational definition of the word trivial is that the system has a fully gapped excitation spectrum and a unique ground state. A more microscopic definition would be that the ground state of the system is a direct product of the individual state spaces of the fundamental degrees of freedom. Since we do not propose an explicit high energy completion (such as a lattice), we have the license to define the trivial phase as the one in which sign$(m) = +1$. Suppose we study a spatial interface between this system and the one in which sign$(m) = -1$ and conclude that there are massless particles localized to the interface. By definition, these degrees of freedom do not belong to the sign$(m) = +1$ system, so we must associate them with the boundary of the system with sign$(m) = -1$. The sign$(m) = -1$ system therefore represents some highly nontrivial entangled state and is said to be the ``non-trivial gapped phase" or the ``topological phase" in condensed matter theory.}
\\\\
Now start in the topological phase ($m < 0$) and turn on interactions. If we can tune the parameter $m$ through $m = 0$ into the phase $m > 0$ \textit{without} closing the bulk gap, then the topological phase is in the same phase as the trivial phase. This means the fermions living on the $(3+1)$-dimensional boundary of the topological phase must have decoupled from the low energy theory: the SM fermions must have obtained mass without breaking $G_{\text{PS}}$ and hence without breaking the electroweak gauge group.
\\\\
Y. Z. You and C. Xu explained that a suitable interaction does in fact exist in order to make this happen \cite{cenke GUT}. Recall the Higgs field introduced in Eq.~(\ref{eq:4-vector higgs}). If $\bra\phi_m\ket = v\,\del_{m4}$, then $SU(2)_L\x SU(2)_R = Spin(4)$ is broken to the diagonal subgroup $SU(2) = Spin(3)$, and all fermions living on the $(3+1)$-dimensional boundary obtain mass at the cost of breaking the $Spin(4)$ part of $G_{\text{PS}}$, which contains the electroweak gauge group.
\\\\
The ground state manifold for the condensed Higgs field is
\begin{equation}
\M_\phi = \frac{Spin(4)}{Spin(3)} = S^3\;.
\end{equation}
In the $(4+1)$-dimensional bulk, spatial infinity is topologically $S^3$:
\begin{equation}
x^m_{\infty} = R\left(\sin\psi \sin\ta\cos\ph,\sin\psi\sin\ta\sin\ph,\sin\psi\cos\ta,\cos\psi \right)\;,\;\; \sum_{m\,=\,1}^4(x_\infty^m)^2 = R^2\;.
\end{equation}
Since
\begin{equation}
\Pi_3(\M_\phi) = Z\;,
\end{equation}
this theory contains pointlike topological defects, called ``hedgehogs" \footnote{These are sometimes also called ``monopoles" in the condensed matter literature, but we will not use this terminology here.}.
\\\\
Let $\phi_{\text{defect}}$ formally be the field operator which creates and annihilates the hedgehogs of $\phi_m$. Since $\bra\phi_m\ket \neq 0$ breaks $Spin(4)$ and gives mass to all of the fermions, we can ask whether condensing the \textit{defect} operator, 
\begin{equation}
\bra\phi_{\text{defect}}\ket \neq 0\;,
\end{equation}
can restore $Spin(4)$ invariance without closing the single particle gap.
\\\\
The main point is this: the fully gapped symmetric phase without intrinsic topological order can be restored by the condensation of a topological defect if and only if the core of the defect has a fully gapped and nondegenerate energy spectrum.
\subsection{Defect core: $n_f = 8k$ Majorana operators}
In 4+1 dimensions, a single 4-component irreducible \textit{Dirac} spinor (or two 4-component Majorana spinors) quantized on a hedgehog background will result in a single 1-component \textit{real} fermion zero mode (``Majorana operator") localized to the core of the defect. This means eight 4-component Dirac spinors (or 16 4-component Majorana spinors) will result in eight 1-component real fermion zero modes at the defect core. So to determine the nature of the hedgehog core, we have to consider eight Majorana operators living in $(0+1)$-dimensional spacetime. 
\\\\
Kitaev and Fidkowski \cite{KF, KF2} explained that in $0+1$ spacetime dimensions, when the number of Majorana fermions is $n_f = 8k$, $k\in Z$, it is possible for interactions to result in a fully gapped and nondegenerate energy spectrum without breaking the symmetry that forbids quadratic terms in the fermion Hamiltonian. This is the key point of their paper, and it is the pioneering behind of the entire field of ``interaction-reduced classification of symmetry protected topological phases." The symmetry of the interaction they proposed can be as large as $Spin(7)$.\footnote{This $Spin(7)$ group is the one which leaves fixed a component of the $8^+$ spinor in $Spin(8)$. The interested reader may wish to review the representation theory of $Spin(2n)$ at this point \cite{wilczek zee}. From the perspective of condensed matter theory, the utility of the $Spin(7)$ and $Spin(8)$ symmetry is simply to find a point in the phase diagram which is most easily amenable to theoretical study. From the perspective of high energy theory, the system of interest is the one with the maximal possible symmetry. This is why, in this context, we work with the $Spin(7)$-covariant notation.}
\\\\
We will use Kitaev's notation for the Majorana operators \cite{kitaev chain}: 
\begin{equation}
c_1,...,c_8\;,\;\; c_a^\da = c_a \;,\;\; \{c_a,c_b\} = \del_{ab}\;.
\end{equation}
These will transform as an irreducible Dirac spinor of an internal $Spin(7)$ symmetry:
\begin{equation}
c_a \sim 8\;\;\text{ of }\;\;Spin(7)\;.
\end{equation}
An antiunitary time reversal transformation which squares to $+1$ can be defined:
\begin{equation}
T:\qquad c_a \to c_a\;,\;\; i \to -i \implies ic_ac_b \to -ic_ac_b\;.
\end{equation}
This forbids all fermion bilinears in the Hamiltonian, so the free fermion Hamiltonian is identically zero:
\begin{equation}
H_{\text{free}} = 0\;.
\end{equation}
Since it takes two Majorana operators to make a physical fermionic oscillator, 
\begin{equation}
a_k \equiv c_{2k-1}+ic_{2k}\;,
\end{equation}
there are four zero modes which can be occupied or not. This means the ground state has degeneracy $2^4 = 16$.
\\\\
To write down the interaction, we will define the eight gamma matrices of $Spin(8)$:
\begin{equation}
\hat\g^a\hat\g^b+\hat\g^b\hat\g^a = \del^{ab}\hat 1\;,\;\; a,b = 1,...,8\;.
\end{equation}
In contrast to the rest of the paper, in this case we express the gamma matrices as abstract operators (denoted by hats) which act in a space of spinors denoted by variables $\e_i = \pm 1$:
\begin{equation}
|\e_1\e_2\e_3\e_4\ket \;,\;\; \e_i = \pm 1\;.
\end{equation}
Those states for which $\prod_{i\,=\,1}^4\e_i = +1$ belong to the $8^+$ representation, and those states for which $\prod_{i\,=\,1}^4\e_i = -1$ belong to the $8^-$ representation. Let 
\begin{equation}
|\psi\ket \equiv \sq(|++++\ket - |----\ket)
\end{equation}
denote a particular linear combination of states in the $8^+$. Then the interaction
\begin{equation}
H_{\text{int}} = g\,\bra \psi|\hat \g^{[a}\hat \g^b\hat \g^c\hat\g^{d]}|\psi\ket c_a c_b c_c c_d
\end{equation}
is manifestly invariant under time reversal $T$ and under the $Spin(7)$ subgroup which leaves the direction $|\psi\ket$ fixed in the $8^+$. It gives a nonzero energy cost to all states (i.e. the Hamiltonian is no longer identically zero), and it singles out a unique state of lowest energy:
\begin{equation}\label{eq:ground state}
|\text{ground state}\ket \equiv \sq(1-a_1^\da a_2^\da a_3^\da a_4^\da)|0\ket 
\end{equation}
where $|0\ket$ is the Fock space vacuum. Recall that the Higgs field $\phi_m \sim (1,2,2)$ is invariant under $Spin(6)$. Therefore, the gapped theory we are considering has a symmetry $Spin(6)$, which is the obvious subgroup of $Spin(7)$ defined by $7 \to 6\oplus 1$. The spinor representation breaks up as
\begin{equation}
8 \to 4\oplus 4^*\;.
\end{equation} 
Therefore, we can think of the complex fermion operators $a_k$ as transforming as the 4-representation of $Spin(6) = SU(4)$:
\begin{equation}
a_k = (a_1,...,a_4) \sim 4\;\;\text{ of }\;\; Spin(6)\;.
\end{equation}
Since $SU(4)$ has the invariant symbol $\e_{ijk\ell}$, the states
\begin{equation}
|0\ket\qquad\text{ and }\qquad \frac{1}{4!}\e_{ijk\ell}\, a^{\da i} a^{\da j} a^{\da k} a^{\da \ell}|0\ket
\end{equation}
are both invariant under $SU(4)$. The linear combination in Eq.~(\ref{eq:ground state}) is therefore invariant under $SU(4)$, and a nonzero expectation value $\bra\phi_{\text{defect}}\ket \neq 0$ would not break $Spin(6)$. Since the defect core is fully gapped, nondegenerate, and invariant under $Spin(6)$, the $Spin(4)$ symmetry can be restored without breaking $Spin(6)$ and without closing the bulk gap. 
\\\\
Therefore, the fermions in the Pati-Salam model (and hence in the SM) can have an interaction-induced mass without breaking the electroweak gauge group. We refer the reader to the original paper \cite{cenke GUT} for a discussion of the dynamical plausibility of the condensation of hedgehogs in this argument. 
\subsection{Defect condensation in $Spin(10)$}\label{sec:defects in 126}
We just discussed the method of defect condensation in the context of restoring the Pati-Salam symmetry $Spin(6)\x Spin(4)$. The absence of defects in the $Spin(10)$ theory occurred because we considered only the 10-vector Higgs field. As mentioned in Eq.~(\ref{eq:so(10) yukawas}), we could also couple the fermions to a 126 representation. 
\\\\
If this field admits topologically nontrivial configurations, then one might expect that the method of defect condensation can be employed to restore the full $Spin(10)$ symmetry. While this field does in fact admit topological defects, unfortunately it seems that condensing those defects cannot restore the fully gapped $Spin(10)$ symmetric phase.
\\\\
The reason is as follows. The minimum for the Higgs potential for a 126-Higgs $\phi_{abcde}$ with negative mass squared gives a vacuum expectation value of the form
\begin{equation}
\bra \phi_{abcde}\ket = v\,\e_{abcde}\;.
\end{equation}
This results in a Majorana mass for the SM-singlet antineutrino and leaves unbroken an $SU(5)$ subgroup of $Spin(10)$. Moreover, this leaves unbroken a $Z_2$ transformation which flips the sign of the spinor representations in $Spin(10)$. This transformation comes from the double cover structure of $Spin(10)$ with respect to $SO(10)$, and it is not contained in the continuous $SU(5)$ subgroup.\footnote{In contrast, this additional structure is not important for the 10-vector since the coset space is $Spin(10)/Spin(9) = SO(10)/SO(9)$.} Therefore,
\begin{equation}
\bra 126\ket:\qquad Spin(10) \to SU(5)\x Z_2\;.
\end{equation}
Because of this $Z_2$, there are vortex solutions labeled by the first homotopy group of the relevant coset space \cite{Z2 vortex GUT, family cloning}:
\begin{equation}
\Pi_1\left( \frac{Spin(10)}{SU(5)\x Z_2}\right) = Z_2\;.
\end{equation}
Thus there are two topologically distinct sectors of Higgs vacuum, one trivial and one nontrivial. Can we condense the operator which creates the nontrivial configurations in order to restore the fully gapped $Spin(10)$ symmetric phase?
\\\\
Unfortunately the answer is no. Since $\bra 126\ket$ leaves $SU(5)$ unbroken, all of the fermions except for the gauge singlet antineutrino do not obtain mass. Thus the Higgs phase of the theory is not fully gapped to begin with, and the condensation of vortex defects cannot restore a fully gapped phase. 
\\\\
This of course does not contradict any of the arguments presented in this paper. All it says is that we do not know how to apply the method of defect condensation to provide an independent argument in support of ``fermion mass without mass terms" in the full $Spin(10)$ invariant theory.
\section{$\Ta$ term and the WZW action}\label{sec:wzw}
As discussed in the main text, for a collection of scalar fields $\phi$ with ground state manifold $\M$ in $(D+1)$-dimensional spacetime, there are no stable topological defects if $\Pi_k(\M) = 0$ for $k = 0,1,...,D$. However, if $\Pi_{D+1}(\M) \neq 0$, there is another type of obstruction related to the Wess-Zumino-Witten (WZW) action. In the present case where $D = 3$ and $\M = S^9$, this obstruction also vanishes:
\begin{equation}
\Pi_4(S^9) = 0\;.
\end{equation}
It is worth explaining the relevance of $\Pi_{D+1}$ in detail, because the process of showing that this obstruction vanishes will also show why 8 flavors of Majorana fermions cannot take advantage of the Kitaev-Wen mechanism, while 16 flavors of Majorana fermions can.\footnote{To a large extent the following will parallel a series of arguments by Y. Z. You and C. Xu \cite{cenke everett SPT} but in a more covariant notation, and these arguments are in turn very closely related to the $\s$-model analysis of Kitaev \cite{kitaev Z16}.} We remind the reader that this is exactly the situation we want: for family unification to work, we require that each ``mirror" family transforming as $16^-$ in $Spin(10)$ decouple from the low energy theory without giving unacceptably large masses to the corresponding $16^+$ fermions and to the electroweak gauge bosons.
\subsection{Eight Weyl fermions: Parent theory}
Split up the $16^-$ fermions into two collections of eight Weyl fermions, each of which transforms as two flavors of $4$-spinor under a $Spin(5)$ flavor symmetry. In order to obtain a covariant notation, it will actually be convenient to begin with a flavor symmetry $Spin(6) = SU(4)$, which we imagine to be broken down to $Spin(5) = USp(4)$ explicitly (not spontaneously). 
\\\\
Let $\nu$ denote a collection of eight left-handed Weyl fermions that transform as the $4^+ \oplus 4^-$ reducible Dirac spinor of $Spin(6)$:
\begin{equation}
\nu = \nu_+ + \nu_-\;,\;\; \nu_\pm \sim 4^\pm\;\;\text{ of }\;\;Spin(6)\;.
\end{equation}
Let $\{\Gs^A\}_{A\,=\,1}^6$ denote the $8\x 8$ gamma matrices of $Spin(6)$, and let $\Cs = -\Gs^2\Gs^4\Gs^6$ denote the corresponding charge conjugation matrix. There are two nonzero fermion bilinears which mix $\nu_+$ and $\nu_-$:
\begin{equation}
\nu_-^T i\s_2\ox \Cs \nu_+ \sim 1\;,\;\; \nu_-^T i\s_2\ox \Cs\Gs^{[A}\Gs^{B]}\nu_+ \sim 15_{\text{adj}}\;.
\end{equation}
We can break $Spin(6)$ to $Spin(5)$ by holding fixed the $6^{\text{th}}$ component of the vector representation: 
\begin{equation}
6\to 5\oplus 1\;.
\end{equation}
Under this decomposition, the two chiral spinors of $Spin(6)$ become the same pseudoreal representation of $Spin(5)$:
\begin{equation}
4^+ \to 4\;,\;\; 4^- \to 4\;.
\end{equation}
The adjoint of $Spin(6)$ breaks up into an adjoint and vector of $Spin(5)$:
\begin{equation}
15_{\text{adj}} \to 10_{\text{adj}}\oplus 5\;.
\end{equation}
Let $\{G^a\}_{a\,=\,1}^5$ denote the $4\x 4$ gamma matrices for $Spin(5)$ and let\footnote{We expect that the reader will be able to discern between the symbol $C$ used here, which denotes a $4\x 4$ matrix, and the $C$ matrix for $Spin(10)$, which is a $32\x 32$ matrix.} $C = -G^2G^4$ denote the corresponding charge conjugation matrix. We will choose the following basis\footnote{The $``\equiv"$ defines our convention for the direct product notation.} for the $Spin(6)$ gamma matrices ($a = 1,...,5$):
\begin{equation}
\Gs^{a} = G^{a}\ox \tau_1 \equiv \ml 0&G^{a}\\ G^{a}&0 \mr\;,\;\;\Gs^6 = I\ox \tau_2 \equiv \ml 0&-iI_{4\x 4}\\ +iI_{4\x 4}&0 \mr\;.
\end{equation}
The $Spin(6)$ charge conjugation matrix is then $\Cs = C\ox \tau_2$, and we have $\Cs\Gs^{a}\Gs^6 = -CG^{a}\ox \tau_1$. If we choose a convention in which $G^5 = G^1G^2G^3G^4$, then the $Spin(6)$ chirality matrix $\Gs_F \equiv i\Gs^1\Gs^2\Gs^3\Gs^4\Gs^5\Gs^6$ is
\begin{equation}
\Gs_F = I_{4\x 4}\ox (-\tau_3) = \ml -I_{4\x 4}&0\\ 0&+I_{4\x 4} \mr\;.
\end{equation}
Let us define 
\begin{equation}
\nu \equiv \ml \nu^1\\ \nu^2 \mr
\end{equation}
so that the chiral spinors are $\nu_+ \equiv \half(I_{8\x 8}+\Gs_F) = \ml 0\\ \nu^2 \mr$ and $\nu_- \equiv \half(I_{8\x 8}-\Gs_F) = \ml \nu^1\\ 0 \mr\;$. In this notation, each $\nu^{\,i}$ transforms as an irreducible 4-spinor of $Spin(5)$, and the label $i = 1,2$ should be thought of as a flavor index. Let us introduce a collection of 5 real scalar fields which transform as a vector under $Spin(5)$ transformations: 
\begin{equation}\label{eq:5-vector}
\phi_a = (\phi_1,...,\phi_5) \sim 5\;\;\text{ of }\;\; Spin(5)\;.
\end{equation}
Then we will find\footnote{To verify this, note that $C\G^{1,...,4}$ are all symmetric matrices, while $C\G^5$ is antisymmetric. For the interested reader, we will show this explicitly in the following non-standard basis (which we will call the ``symplectic basis"): 
\begin{equation*}
G^{1,2,3} = \tau_2\ox \tau_{1,2,3}\;,\;\; G^4 = \tau_1\ox I\;,\;\; G^5 =  G^1 G^2 G^3 G^4 = \tau_3\ox I\;.
\end{equation*}
The charge conjugation matrix is $C = -G^2G^4 = \tau_3\ox i\tau_2$, so the first four matrices are:
\begin{equation*}
CG^{1,2,3} = \tau_1\ox \tau_2\tau_{1,2,3}\;,\;\; CG^4 = i\tau_2\ox i\tau_2\;.
\end{equation*}
These matrices are symmetric. However, the fifth matrix is antisymmetric:
\begin{equation*}
CG^5 = I\ox i\tau_2 = \ml 0&I\\ -I&0 \mr \equiv J
\end{equation*}
where we have identified the invariant symbol $J$ of $USp(4)$.
}:
\begin{align}\label{eq:coupling to 5-vector from so(6)}
\phi_a\, \nu_+^T &i\s_2\ox \Cs\Gs^a\Gs^6 \nu_-  =\nonumber\\
& -\half \sum_{a\,=\,1}^4 \phi_a\,\nu^{T\,i} i\s_2\ox CG^a \ml 0&1\\ 1&0 \mr_{ij}\nu^{\,j}+\half \phi_5\,\nu^{T\,i} i\s_2\ox CG^5 \ml 0&1\\ -1&0 \mr_{ij} \nu^{\,j}\;.
\end{align}
The Lagrangian for 8 fermions without a mass term but with a Yukawa interaction of the form in Eq.~(\ref{eq:coupling to 5-vector from so(6)}) will be the physical theory of interest.
\\\\
It will also be very useful to introduce a mass term for this theory. This will come from the $Spin(6)$ singlet fermion bilinear, which in this notation becomes:
\begin{equation}\label{eq:so(6) singlet}
\nu_+^T i\s_2\ox \Cs\nu_- = \half \nu^{T\,i} i\s_2\ox C (\tau_2)_{ij}\,\nu^{\,j}\;.
\end{equation}
The overall sign of the mass parameter will be captured by a sixth scalar field,
\begin{equation}\label{eq:non-dynamical field}
\ph \sim 1\;\;\text{ of }\;\; Spin(5)\;,
\end{equation}
which couples to the fermion bilinear in Eq.~(\ref{eq:so(6) singlet}). In the physical theory of interest, we should think of the $\phi_a$ in Eq.~(\ref{eq:5-vector}) as dynamical quantum fields, but we should think of the $\ph$ defined in Eq.~(\ref{eq:non-dynamical field}) as a fixed constant background field.
\\\\
In order to determine whether the interacting theory of 8 fermions coupled to the $\phi_a$ as in Eq.~(\ref{eq:coupling to 5-vector from so(6)}) is massive or massless when $\ph = 0$, we will need a smooth interpolation between a free theory with $\ph = +1$ and a free theory with $\ph = -1$. This interpolating path will be obtained by defining a collection of six dynamical scalar fields,
\begin{equation}\label{eq:six fields}
\Phi_A = (\Phi_0,\Phi_1,...,\Phi_5)\;.
\end{equation}
Define the following $8\x 8$ matrices:
\begin{equation}
M^0_{ij} \equiv C\,(\tau_2)_{ij}\;,\;\;M^{1,...,4}_{ij} \equiv CG^{1,...,4}\, (\tau_1)_{ij}\;,\;\; M^5_{ij} \equiv CG^5\,(-i\tau_2)_{ij}\;.
\end{equation}
We will consider the following Lagrangian:
\begin{equation}\label{eq:parent lagrangian}
\la = \sum_{i\,=\,1}^2\nu^{\da\, i}\bar \s^\mu \,i\pa_\mu \nu^{\,i}-m\left[\nu^{T\,i}i\s_2\ox \left( M^0_{ij}\,\Phi_0 + \sum_{a\,=\,1}^5M^a_{ij}\,\Phi_a\right)\nu^{\,j} +h.c.\right]\;.
\end{equation}
This will serve as a sort of ``parent" theory for the arguments that follow. 
\subsection{Eight Weyl fermions: Trivial vs. Topological}
The Lagrangian in Eq.~(\ref{eq:parent lagrangian}) serves as a parent theory in the following sense. The models we are actually interested in are obtained by imposing the constraint
\begin{equation}
\Phi_0^{\,2}+\sum_{a\,=\,1}^5\Phi_a^{\,2} = 1
\end{equation}
and considering various points on this $S^5$ in field space. First consider the North pole:
\begin{equation}
\Phi_0 = +1\;,\;\; \Phi_a = 0\;.
\end{equation}
This corresponds to a theory of eight free fermions with Majorana mass terms:
\begin{equation}\label{eq:trivial phase}
\la_{\text{trivial}} = \sum_{i\,=\,1}^2\nu^{\da\,i}\bar\s^\mu\,i\pa_\mu\nu^{\,i}-m\,\e_{ij}\left( \nu^{T\,i}\s_2\ox C\,\,\nu^{\,j}+h.c.\right)\;.
\end{equation}
Next consider the South pole:
\begin{equation}
\Phi_0 = -1\;,\;\;\Phi_a = 0\;.
\end{equation}
Naively this also corresponds to a theory of eight free fermions with Majorana mass terms:
\begin{equation}\label{eq:topological phase}
\la_{\text{topological}} = \sum_{i\,=\,1}^2\nu^{\da\,i}\bar\s^\mu\,i\pa_\mu\nu^{\,i}+m\,\e_{ij}\left( \nu^{T\,i}\s_2\ox C\,\,\nu^{\,j}+h.c.\right)\;.
\end{equation}
However, as recent developments in condensed matter theory tell us, a crucial observation is that the sign of the fermion mass term is important and cannot simply be compensated by a field redefinition. If Eq.~(\ref{eq:trivial phase}) describes an ordinary free fermion theory, then on a spacetime with spatial boundaries the theory described by Eq.~(\ref{eq:topological phase}) will have massless particles living on the $(2+1)$-dimensional boundary (see Sec.~\ref{sec:2pi}).
\\\\
In condensed matter language, the Lagrangian of Eq.~(\ref{eq:trivial phase}) is said to describe a ``trivial" superconductor, while the Lagrangian of Eq.~(\ref{eq:topological phase}) is said to describe a ``topological" superconductor. Both models have the same fully gapped bulk spectra, but one model has a gapless boundary and hence describes a highly nontrivial entangled state.
\\\\
To a particle physicist, the statement that the sign of the fermion mass term cannot simply be compensated by a field redefinition may feel rather foreign. The reader should remember that the same low energy effective field theory can arise from distinct high energy completions, and a naive treatment of the effective field theory may not completely capture all important aspects of the underlying high energy completion. 
\\\\
In the context of topological superconductors, the pertinent high energy completion is a lattice regularization: the simplest example of this is the $(1+1)$-dimensional Kitaev chain \cite{kitaev chain}. In the present context, the reader may consider the appropriate high energy regularization to be the parent theory in Eq.~(\ref{eq:parent lagrangian}) whose $\s$-model has a target space $S^5$. The important effects \textit{can} be captured by a quantum field theory without committing to a lattice regularization, but that quantum field theory will have some perhaps unfamiliar properties (such as an extra spatial dimension or a scalar field with a domain wall profile) which may not be immediately obvious from the perspective of local observables in low energy physics.
\\\\
Finally, the third configuration to consider is\footnote{The notation in Eq.~(\ref{eq:equator configuration}) may appear redundant, so let us explain it. We use this notation to emphasize that the physical theory of interest is the one in which the $\phi_a$ in Eq.~(\ref{eq:5-vector}) couple to the fermions as in Eq.~(\ref{eq:coupling to 5-vector from so(6)}) with a mass parameter $\ph$. The $\Phi_A$ in Eq.~(\ref{eq:parent lagrangian}) are auxiliary dynamical fields which are to be fixed in some way to reduce the parent Lagrangian to the physical theory of interest. The $\phi_a$ are defined only on the ``equator" $S^4$ while the $\Phi_A$ are defined on the entire $S^5$.} the ``equator" $S^4$:
\begin{equation}\label{eq:equator configuration}
\Phi_0 = 0\;,\;\; \Phi_a = \phi_a\;.
\end{equation}
This corresponds to a theory of eight fermions without mass terms interacting with a 5-component dynamical scalar field:
\begin{equation}\label{eq:yukawa for eight fermions}
\la_{\text{Yuk}} = \sum_{i\,=\,1}^2\nu^{\da\,i}\bar\s^\mu\,i\pa_\mu\nu^{\,i}-m\,\sum_{a\,=\,1}^5\phi_a\left( \nu^{T\,i}i\s_2\ox M^a_{ij}\,\nu^{\,j}+h.c.\right)\;.
\end{equation}
Eq.~(\ref{eq:yukawa for eight fermions}) describes fermions coupled to a $\s$-model with target space $S^4$. In contrast, the five fields $\Phi_a$ satisfy $\sum_{a\,=\,1}^5\Phi_a^{\,2} = 1-\Phi_0^2$, and Eq.~(\ref{eq:parent lagrangian}) describes fermions coupled to a $\s$-model with target space $S^5$.
\\\\
At weak coupling $(m \ll 1)$, the Lagrangian in Eq.~(\ref{eq:yukawa for eight fermions}) describes massless fermions interacting with a Higgs field, as usual. The question is whether the fermions remain massless even when the Yukawa coupling $m$ is of intermediate strength (not too weak for perturbation theory to apply, and not too strong for perturbation theory in $1/m$ to apply).
\\\\
As we said before, for a fixed value of $\sum_{a\,=\,1}^5\phi_a^{\,2}$, we have $\phi_a\in S^4$. This means $\Pi_k(S^4) = 0$ for $k = 0,1,2,3$, and therefore there are no topological defects. But we now have:
\begin{equation}
\Pi_4(S^4) = Z\;.
\end{equation}
The physical meaning of this is that, after performing the path integral over the fermions, the effective action for $\phi_a$ may contain a theta term (we will use Euclidean signature):
\begin{equation}
S_\Ta[\phi] = i\Ta\int_{S^4}\frac{1}{\W_4}\e^{abcde}\phi_a\, d\phi_b\wedge d\phi_c \wedge d\phi_d \wedge d\phi_e\;.
\end{equation} 
The question now is: what is the correct value of $\Ta$?
\subsection{Sigma model with $\Ta = 2\pi$}\label{sec:2pi}
To figure this out, let us work in compactified Euclidean spacetime (topologically $S^4$) and return to the parent Lagrangian in Eq.~(\ref{eq:parent lagrangian}). We will follow closely the calculations of Abanov and Wiegmann \cite{abanov wiegmann}.
\\\\
Let $V \equiv \sum_{A\,=\,0}^5M^A\Phi_A$. Then we can integrate out the fermions and obtain an effective action $S_{\text{eff}}[V] = \half i\int d^4x \bra x| \tr\ln(i\!\!\not\!\pa-m V)|x\ket$. The imaginary part of $\del S_{\text{eff}}[V] \equiv S_{\text{eff}}[V+\del V]-S_{\text{eff}}[V]$ is:
\begin{equation}
\im(\del S_{\text{eff}}[V]) \propto \e^{ABCDEF}\e^{\mu\nu\rho\s}\Phi_A \pa_\mu \Phi_B \pa_\nu \Phi_C \pa_\rho \Phi_D \pa_\s\Phi_E\del\Phi_F\;.
\end{equation}
To restore $\im(S_{\text{eff}})$ from its variation \cite{witten current algebra}, introduce a parameter $u \in [0,1]$ and define an extension $\widetilde\Phi_A(x,u)$ of the scalar fields into the unit ball $\Bs^5 = S^4\x [0,1]$:
\begin{equation}
\widetilde\Phi_A(x,0) = \del_{A0}\;,\;\; \widetilde\Phi_A(x,1) = \Phi_A(x)\;;\;\;A = 0,1,...,5\;.
\end{equation}
The imaginary part of the effective action is then the WZW action at level $k = 1$:
\begin{equation}\label{eq:wzw}
\im(S_{\text{eff}}[\Phi]) = \frac{2\pi}{5!\;\W_5}\int_{\Bs^5}\e^{ABCDEF}\widetilde\Phi_A d\widetilde\Phi_B\wedge d\widetilde\Phi_C \wedge d\widetilde\Phi_D \wedge d\widetilde\Phi_E\wedge d\widetilde\Phi_F\;.
\end{equation}
As we mentioned before, this model can be reduced to a theory of massive fermions interacting with the five fields $\phi_a(x)$ by constraining the six fields $\widetilde\Phi_A(x,u)$ to a target space $S^5$. Impose the constraint
\begin{equation}
\widetilde\Phi_0^{\,2}+\sum_{a\,=\,1}^5\widetilde\Phi_a^{\,2} = 1
\end{equation}
and consider the following field configuration:
\begin{equation}\label{eq:config for theta}
\widetilde\Phi_0(u) = \cos(\al(u))\;,\;\; \widetilde\Phi_a(x,u) = \phi_a(x)\,\sin(\al(u))\;,\;\; \al(0) = 0\;,\;\;\al(1) = \beta\;.
\end{equation}
Here $\al(u)$ is only a function of $u\in[0,1]$, $\beta$ is a constant, and the $\phi_a(x)$ are constrained to the surface of a unit $S^4$ as before. If we plug Eq.~(\ref{eq:config for theta}) into Eq.~(\ref{eq:wzw}), we will find \cite{abanov wiegmann}:
\begin{equation}\label{eq:theta term}
\im(S_{\text{eff}}) = \Ta(\beta)\;\int_{S^{4}}\frac{1}{\W_4}\, \e^{a_1...a_5} \phi_{a_1} d\phi_{a_2}\wedge ... \wedge d\phi_{a_5}\;,\;\; \Ta(\beta) = 2\pi \;\frac{\int_0^\beta d\al\; \sin^{4}\al}{\int_0^\pi d\al\; \sin^{4}\al}\;.
\end{equation}
This is the $\Ta$ term for the $\s$-model with target space $S^{4}$. It computes the degree of the map $\phi:S^4 \to S^4$.
\\\\
Recall the original Yukawa interaction in Eq.~(\ref{eq:parent lagrangian}). The expression in Eq.~(\ref{eq:theta term}) tells us that if we interpolate continuously from a free fermion theory with mass parameter $m_{\text{eff}} = +m$ (i.e. $\beta = 0$) to a free fermion theory with mass parameter $m_{\text{eff}} = -m$ (i.e. $\beta = \pi$), we necessarily pick up a theta term with parameter $\Ta = 2\pi$. A simple field redefinition to compensate for the change in sign of the fermion mass would miss this important contribution to the path integral.
\\\\
The critical point (namely the equator, $\Phi_0 = 0$) exhibits an enhanced symmetry
\begin{equation}
Z_2:\qquad \nu^{T\,i}i\s_2 \nu^{\,j} \to -\nu^{T\,i}i\s_2 \nu^{\,j}\;,\;\; \phi_a \to -\phi_a\;.
\end{equation}
This unitary $Z_2$ transformation commutes with the $Spin(5)$ flavor symmetry. As long as $Spin(5)\x Z_2$ remains unbroken, no fermion mass term can appear in the Lagrangian at any order in perturbation theory.
\subsection{Interface between $\Ta = 2\pi$ and $\Ta = 0$}\label{sec:interface}
Let us go back to Eq.~(\ref{eq:wzw}). We would like to consider a $(2+1)$-dimensional interface between two $(3+1)$-dimensional phases, one with $m_{\text{eff}} = +m$ and the other with $m_{\text{eff}} = -m$. For this purpose, instead of the parametrization in Eq.~(\ref{eq:config for theta}), we will consider the following profile \cite{vishwanath senthil}:
\begin{equation}\label{eq:config for interface}
\widetilde\Phi_0(x) = \cos(\al(z))\;,\;\; \widetilde\Phi_a(x,u) = f_a(x,u)\;\sin(\al(z))\;,\;\; \al(z) = \pi \ta(z) = \left\{ \begin{matrix} 0\;,\;\; z < 0\\ \frac{\pi}{2}\;,\;\; z = 0\\ \pi\;,\;\; z > 0 \end{matrix} \right.
\end{equation}
with $\sum_{a\,=\,1}^5f_a^2 = 1$. From the previous section, we see that this gives us a $(3+1)$-dimensional $\s$-model with $\Ta = 0$ in the region $z < 0$ and a $(3+1)$-dimensional $\s$-model with $\Ta = 2\pi$ in the region $z > 0$. The goal is to study the $(2+1)$-dimensional interface located at $z = 0$. Since the phase with $\Ta = 0$ is a trivial gapped phase, the possibly nontrivial degrees of freedom at $z = 0$ should be thought of as the boundary of the phase with $\Ta = 2\pi$.
\\\\
If we plug Eq.~(\ref{eq:config for interface}) into Eq.~(\ref{eq:wzw}), we will obtain a level 1 WZW term at $z = 0$:
\begin{equation}\label{eq:wzw at interface}
S_{\text{WZW}}[n] = i\frac{2\pi }{4!\;\W_4} \int_{\Bs^4} \e^{a_1... a_5}\; N_{a_1} dN_{a_2}\wedge ... \wedge dN_{a_5}\;,
\end{equation}
where
\begin{equation}
N_a(t,x,y,u) \equiv f_a(t,x,y,0,u)\;,\;\; N_a(t,x,y,0) = \del_{a,5}\;,\;\; N_a(t,x,y,1) = n_a(t,x,y)\;.
\end{equation}
Therefore, the interface between the $\s$-model with $\Ta = 0$ and the $\s$-model with $\Ta = 2\pi $ is described by a $\s$-model with a target space $S^4$ and a WZW term at level $k = 1$:
\begin{equation}\label{eq:interface}
S_{\text{interface}}[n] = \int_{S^3} d^{2+1}x\,\frac{1}{2g}(\pa_\mu n_a)^2+S_{\text{WZW}}[n]\;,\;\; \sum_{a\,=\,1}^5 n_a^2 = 1\;.
\end{equation}
For a $\s$-model with target space $S^2$ in $0+1$ dimensions, the addition of a level 1 WZW term results in a twofold degenerate ground state. For a $\s$-model with target space $S^3$ in $1+1$ dimensions, the addition of a level 1 WZW term results in a conformal field theory (this is the statement of nonabelian bosonization for the $SU(2)$ model). In $2+1$ dimensions the analysis is more difficult, but the addition of the level 1 WZW term to the model in Eq.~(\ref{eq:interface}) is also expected to result in a conformal field theory \cite{eg 1, eg 2}.
\\\\
Since this interface is the boundary of the $(3+1)$-dimensional $\s$-model with target space $S^4$ and theta parameter $\Ta = 2\pi$, we conclude that, despite the seemingly innocuous free fermion Lagrangian in Eq.~(\ref{eq:topological phase}), this theory is in fact gapless on a space with boundaries. In condensed matter language, it is a ``topological" superconductor rather than a ``trivial" superconductor.
\\\\
There is a quantum phase transition between the trivial and topological phases, and hence the Yukawa theory in Eq.~(\ref{eq:yukawa for eight fermions}) is massless. It is not possible for fermions to have mass without mass terms for $n_f = 8$ flavors in $3+1$ dimensions. This is what is meant by the ``WZW obstruction" to realizing the symmetric gapped phase.
\subsection{16 Weyl fermions}
We argued that 8 Weyl fermions coupled to 5 scalar fields cannot obtain masses without mass terms. But $Spin(10)$ unification gives us 16 Weyl fermions coupled to 10 scalar fields. So in fact we have two copies of the system studied in the previous section. 
\\\\
Let us work with the effective bosonic theory and denote the two collections of scalar fields as follows:
\begin{equation}
\phi_a^{\,(I)} = (\phi_1^{\,(I)},\,...\,,\,\phi_5^{\,(I)})\;,\;\; I = 1,2\;.
\end{equation} 
We can turn on interaction between these two bosons which is invariant under $Spin(5)\x Z_2$ transformations:
\begin{equation}
\la_{\text{int}} = \As\,\sum_{a\,=\,1}^5\phi_a^{\,(1)}\phi_a^{\,(2)}\;.
\end{equation}
When $\As$ is large, the two scalar fields prefer to be aligned in $Spin(5)$. Therefore we should do perturbation theory about the configuration
\begin{equation}
\phi_a^{\,(1)} = \phi_a^{\,(2)} \equiv \phi_a^{\,(0)}\;.
\end{equation}
The leading order theory is therefore a sigma model for $\phi_a^{\,(0)}$ with effective $\Ta$ parameter
\begin{equation}
\Ta_{\text{eff}}^{\,(0)} = \Ta^{\,(1)}+\Ta^{\,(2)} = 4\pi\;.
\end{equation}
So when the coupling $\As$ is large, we can approximate the theory of 16 interacting Weyl fermions as two collections of 8 Weyl fermions coupled to a single $\s$-model with target space $S^4$ described by the field $\phi_a^{\,(0)}$. 
\\\\
The $\s$-model with $\Ta = 4\pi$ is smoothly connected to the trivial gapped phase with $\Ta = 0$ \cite{cenke grassmannian}. We can then repeat the argument of Sec.~\ref{sec:interface}, but this time with $\Ta = 4\pi$ in the region $x^3 > 0$ (i.e. we can use $\al(x^3) = 2\pi\ta(x^3)$ in the notation of that section). There will be no WZW term at the interface and therefore no massless degrees of freedom at the boundary of the phase on the right. 
\\\\
This tells us that it is possible to continuously deform between the phase with $m_{\text{eff}}^{\,(1)} = m_{\text{eff}}^{\,(2)} = +m$ and the phase with $m_{\text{eff}}^{\,(1)} = m_{\text{eff}}^{\,(2)} = -m$ without passing through a point in the phase diagram which is either gapless or spontaneously breaks the symmetry which protects the fermion mass term. Therefore, it \textit{is} possible for 16 Weyl fermions in 3+1 dimensions to have mass without explicit mass terms in the Lagrangian. 
\\\\
The continuous part of the global symmetry group in the path that we have discussed is $Spin(5)$, which can be gauged via minimal coupling. As discussed previously, if the gauge coupling is weak, the $Spin(5)$ gauge bosons will remain massless as usual. The question is whether it is possible to enlarge $Spin(5)\x Z_2$ into the $Spin(10)$ symmetry of grand unification, in which the ten bosons would transform as a 10-vector, and the 16 Weyl fermions would transform as a chiral $16^-$. (Remember we are trying to gap out the mirror fermions without breaking the $Spin(10)$ gauge symmetry.) 
\\\\
Since the path we have discussed does not spontaneously break any symmetry, we expect that enlarging the symmetry to $Spin(10)$ without substantially changing the strength of the interactions should not result in a vacuum expectation value for the scalar field. Moreover, if the $Spin(10)$ invariant interaction explicitly breaks all possible anomalous global symmetries in the mirror sector, then no new massless states should appear when enlarging the symmetry from $Spin(5)\x Z_2$. 
\\\\
Therefore, there should be an interacting path which gaps out all mirror fermions without breaking the $Spin(10)$ gauge symmetry, and we arrive at the same conclusion obtained via the Kitaev-Wen argument. 
\section{General formulas for $\Ta$ terms and the WZW action}\label{sec:wzw and theta}
In this appendix we collect some useful formulas for WZW terms in any number of spacetime dimensions. See Abanov and Wiegmann \cite{abanov wiegmann} for a more comprehensive treatment of topological terms in $\s$-models coupled to fermions.
\\\\
Let the number of spacetime dimensions be $d$, and let us work with compactified Euclidean spacetime $S^d$. Introduce an additional parameter $u\in [0,1]$ and define the unit ball $\Bs^{d+1} = S^d\x [0,1]$. 
\\\\
Define a collection of $d+2$ scalar fields which depend on $x^\mu\in S^d$ and $u$:
\begin{equation}
\Phi_A = (\Phi_0,\Phi_1,...,\Phi_{d+1})\;.
\end{equation}
The WZW action at level $k$ is defined as:
\begin{equation}
S_{\text{WZW}}[\Phi] = i\frac{2\pi k}{(d+1)!\;\W_{d+1}}\int_{\Bs^{d+1}}\e^{A_0A_1...A_{d+1}}\Phi_{A_0}d\Phi_{A_1}\wedge ... \wedge d\Phi_{A_{d+1}}
\end{equation}
where
\begin{equation}
\W_{n} \equiv \frac{2\pi^{(n+1)/2}}{\G(\frac{n+1}{2})}\;.
\end{equation}
Let $a = 1,...,d+1$ label the scalar fields $\Phi_1,...,\Phi_{d+1}$, and impose the constraint
\begin{equation}
\Phi_0^{\,2}+\sum_{a\,=\,1}^{d+1}\Phi_a^{\,2} = 1\;.
\end{equation}
Then $\Phi_A\in S^{d+1}$. It will be useful to consider two different profiles for $\Phi_A(x,u)$ subject to this constraint.
\subsection{Profile 1: Reduce WZW to $\Ta$ term}
If
\begin{equation}
\Phi_0(x,u) = \cos(\al(u))\;,\;\; \Phi_a(x,u) = \phi_a(x)\sin(\al(u))\;,\;\; \al(0) = 0\;,\;\; \al(1) = \beta
\end{equation}
then
\begin{equation}
S_{\text{WZW}}[\Phi] = i\left( 2\pi k\;\frac{\int_0^\beta d\al\,\sin^{d} \al}{\int_0^\pi d\al\,\sin^{d}\al}\right)\int_{S^{d}}\frac{1}{\W_{d}}\;\e^{a_1...a_n}\phi_{a_1}d\phi_{a_2}\wedge ... \wedge d\phi_{a_{d+1}}\;.
\end{equation}
To verify this, note that:
\begin{equation}
\frac{\W_n}{\W_{n-1}} = \frac{\pi^{1/2}\G(\frac{n}{2})}{\G(\frac{n+1}{2})} = \int_0^\pi d\al\;\sin^{n-1}\al\;.
\end{equation}
The North pole of the target space $S^{d+1}$ is $\Ta = 0$, the South pole is $\Ta = 2\pi k$, and the equator $S^{d}$ is $\Ta = \pi k$.
\subsection{Profile 2: Domain wall between $\Ta = 0$ and $\Ta = 2\pi$}
If
\begin{equation}
\Phi_0(x,u) = \cos(\al(x^{d-1}))\;,\;\; \Phi_a(x,u) = f_a(x,u)\sin(\al(x^{d-1}))\;,\;\; \al(x^{d-1}) = \pi \ta(x^{d-1})
\end{equation}
then
\begin{equation}
S_{\text{WZW}}[\Phi] = i\frac{2\pi k}{d!\;\W_d} \int_{\Bs^{d}}\e^{a_1...a_{d+1}}N_{a_1} dN_{a_2}\wedge ... \wedge dN_{a_{d+1}}\;,
\end{equation}
where
\begin{equation}
N_a(x^0,...,x^{d-2},u) \equiv f_a(x^0,...,x^{d-2},0,u)\;.
\end{equation}
Therefore:
\begin{equation}
S_{\text{interface}}[n] = \int_{S^{d-1}}\frac{1}{2g}dn_a\wedge *dn_a+i\frac{2\pi\,k}{d!\;\W_d} \int_{\Bs^d}\e^{a_1...a_{d+1}}N_{a_1} dN_{a_2}\wedge ... \wedge dN_{a_{d+1}}\;,
\end{equation}
where
\begin{equation}
N_a(x,0) = \del_{a,d+1}\;,\;\; N_a(x,1) = n_a(x)\;.
\end{equation}
This tells us that the $(d-1)$-dimensional interface between a $d$-dimensional $\s$-model with $\Ta = 0$ and a $d$-dimensional $\s$-model with $\Ta = 2\pi k$ contains a $\s$-model with WZW term at level $k$.

\end{document}